\documentclass[10pt,shortpaper,twoside,web] {ieeecolor}

\usepackage{generic}
\usepackage{cite}
\usepackage{amsmath,amssymb,amsfonts}
\usepackage{algorithmic}
\usepackage{graphicx}
\usepackage{textcomp}
\usepackage{psfrag} 
\usepackage{cuted}
\usepackage{float}
\usepackage{graphicx} 
\usepackage{amsmath}  
\usepackage[fancythm,fancybb,morse,ieee]{jphmacros2e} 
\usepackage{array}
\usepackage{bbm}
\usepackage{mathrsfs}
\usepackage{epstopdf}
\DeclareMathAlphabet{\mathpzc}{OT1}{pzc}{m}{it}
\usepackage{amsmath}
\usepackage[fancythm,fancybb,morse,ieee]{jphmacros2e}
\usepackage{caption}
\usepackage{subcaption}
\usepackage{algorithm}
\usepackage{algorithmic}
\usepackage{multirow}
\usepackage{csquotes}
\usepackage{url}
\newtheorem{thm}{Theorem}
\newtheorem{lem}{Lemma}

\newtheorem{defn}{Definition}

\newtheorem{rem}{Remark}

\def\BibTeX{{\rm B\kern-.05em{\sc i\kern-.025em b}\kern-.08em
		T\kern-.1667em\lower.7ex\hbox{E}\kern-.125emX}}
\begin{document}
	\title{Unifying Direct and Indirect Learning for Safe Control of Linear Systems}
	\author{Amir Modares, Niyousha Ghiasi,  Bahare Kiumarsi, and Hamidreza Modares
		\thanks{ }
		\thanks{A. Modares is an independent researcher (e-mail: amir.modares.81@gmail.com). N. Ghiasi, B. Kiumarsi, and H. Modares are with Michigan State University, USA (emails: ghiasini@msu.edu; kiumarsi@msu.edu; modaresh@msu.edu).}}
	
\maketitle
\thispagestyle{empty} 
\begin{abstract} 
\textcolor{blue}{This paper develops learning-enabled safe controllers for linear systems subject to system uncertainties and bounded disturbances. Given the disturbance zonotope, the data-based closed-loop dynamics (CLDs) are first characterized using a matrix zonotope (MZ), and refined through several steps to yield a constrained matrix zonotope (CMZ). This refinement is achieved by introducing conformal equality constraints that eliminate incompatible disturbance realizations. More precisely, prior knowledge and observed data are used separately to construct CMZ representations of disturbance sequences that conform to both data and prior knowledge, and are intersected by the initial MZ of the disturbance sequence, producing a refined CMZ. This approach reduces conservatism. To further reduce the conservativeness, we unify open-loop learning with closed-loop learning by presenting a novel set-membership identification method that models open-loop dynamics as a CMZ. The prior knowledge serves as an initial feasible open-loop model set (FOLMS) of this CMZ, which is refined into a posterior set whenever new informative online data becomes available. This posterior FOLMS then adaptively replaces the prior knowledge set employed in the disturbance elimination of the closed-loop learning process.
The resulting refined parameterized set of CLD is subsequently leveraged to directly and adaptively learn a controller that robustly enforces safety. Toward this goal, we formulate a linear programming problem that guarantees $\lambda$-contractiveness of a polyhedral safe set. A simulation example is provided to validate the effectiveness of the proposed approach and support the theoretical results. }
\end{abstract}

\IEEEpeerreviewmaketitle

\section{Introduction}

\IEEEPARstart{D}{ata}-based control design is categorized into direct data-driven control and indirect data-driven control \cite{Data1}-\cite{DI1}. The former parameterizes the controller and directly learns its parameters to satisfy control specifications. The latter learns a set of system models that explain the data and then synthesizes a robust controller based on this set of models. 

A recent popular direct data-driven approach \textit{characterizes the set of all closed-loop systems} using data \cite{Data2}-\cite{Data6}. The control-oriented nature of this approach offers the potential to improve control system performance~\cite{Data4} and data requirements \cite{Data5}. 
However, despite its advantages, direct learning faces challenges in incorporating available prior knowledge into the learning framework. Effectively leveraging prior model knowledge can significantly enhance the performance of direct learning by reducing conservatism and improving robustness. In particular, it can mitigate sensitivity to noise and disturbances, as demonstrated in~\cite{Ozay} for the linear quadratic regulator problem. 

\textcolor{blue}{Several indirect learning approaches have successfully incorporated prior knowledge into the learning process. This includes set-membership identification methods~\cite{ID1}–\cite{ID2}, where uncertainty is typically represented by ellipsoidal sets, and the S-procedure is used to intersect the data-consistent and the prior knowledge ellipsoids. However, the S-procedure often introduces conservatism due to the inherent overapproximation in ellipsoidal intersection representations. Physics-informed learning-enabled settings ~\cite{PyI1,PyI2,PyI3} rely on the S-procedure or scenario-based methods to incorporate prior knowledge. Prior knowledge of the disturbance models is utilized in \cite{New} to enhance the learning of the disturbance dynamics, while assuming that the system model dynamics are known. Again, ellipsoidal uncertainty sets and S-procedure are leveraged in \cite{New}. An alternative line of research incorporates prior knowledge into open-loop zonotope-based system identification for scalar systems \cite{ID2z} and safety verification \cite{ID3}. The use of zonotope-based closed-loop learning, along with a unified framework that integrates open-loop and closed-loop learning with prior knowledge—rather than relying solely on open-loop learning—and its application to control synthesis, rather than verification, remains largely unsettled.}
\textcolor{blue}{To combine the strengths of direct and indirect learning, this paper integrates them with prior knowledge to develop safe controllers. We first construct a matrix zonotope (MZ) to represent closed-loop models from data, and then refine it into a constrained matrix zonotope (CMZ) by removing disturbances inconsistent with prior knowledge and identified open-loop dynamics. Unlike ellipsoidal sets, CMZs more accurately capture closed-loop uncertainty by enabling precise operations, such as the intersection of two CMZs introduced in this paper. We also introduced a zonotope-based set-membership method to update the feasible open-loop model set (FOLMS) from online data. The FOLMS is initialized with prior knowledge, and its posterior updates inform closed-loop learning and controller synthesis. A data informativeness condition enables an adaptive algorithm that updates the closed-loop model and controller. The resulting linear programming-based approach ensures robust, adaptive, and safe control. Simulation results demonstrate the proposed method’s advantage over conventional direct and indirect learning approaches.} 

\noindent \textbf{Notations and Definitions.} Throughout the paper, $\mathbb{R}^n$ and $\mathbb{R}^{n \times m}$ denote the sets of real-valued vectors and matrices, respectively. $I_p$ is the $p \times p$ identity matrix, $\bar{\mathbf{1}}$ is the column vector of ones, and $B_\infty$ denotes the unit hypercube in $\mathbb{R}^n$. The Kronecker product of $A$ and $B$ is denoted by $A \otimes B$, and $\text{diag}(g_1,\dots,g_n)$ defines a diagonal matrix with entries $g_1$ to $g_n$. For a vector $x \in \mathbb{R}^n$, $x_i$ is its $i$-th element. The zero matrix is denoted by $\mathbf{0}_{n \times m}$.
For a matrix $X$, $X^{\top}$ is its transpose, \textcolor{blue}{$X_{i,:}$ is its $i$-th row, $X_{:,j}$ is its $j$-th column}, and $X_{i,j}$ is its $(i,j)$-th entry. $X^{\bot}$ denotes a matrix whose columns form a basis for the kernel of X. \textcolor{blue}{The element-wise absolute value of $X$ obtained by replacing each element with its absolute value is denoted by $|X|$}, and $\| \cdot \|$ represents the infinity norm of a vector or matrix.
\textcolor{blue}{Given $G = [G_{:,1}, \dots, G_{:,s_1}] \in \mathbb{R}^{n \times ms_1}$ with $G_{:,i} \in \mathbb{R}^{n \times m}$ for \( i = 1, \dots, s_1 \), and $\bar N \in \mathbb{R}^{m \times p}$, we define $G \circ \bar N=G (I_{s_1} \otimes \bar N)$.}
Finally, $\text{Vec}(G) = [\text{Vec}(G_{:,1}), \dots, \text{Vec}(G_{:,s_1})]$, where $\text{Vec}(G_{:,i})$ stacks the matrix $G_{:,i}$ into a column vector.

\textcolor{blue}{\begin{defn} \textbf{\big((Constrained) Zonotope\big) \cite{ID3}} \label{defcz}
Given a generator matrix $G \in \mathbb{R}^{n \times s}$ and a center $c \in \mathbb{R}^n$, a constrained zonotope of dimension $n$  with $s$ generators is represented by 
\begin{align*} 
& \cal{C}\!=\!\big<G,c,\bar A^c,\bar b^c\big>  \!=\!\Big\{\!x \in \mathbb{R}^n\!\!:\! x\!=\!G \, \zeta\! +\!c, \, \big\Vert \zeta \big\Vert\! \leq \!1, \,\,\,  \!\!\bar A^c \zeta\!=\!\bar b^c\! \Big\},
\end{align*}
where $\bar A^c \in \mathbb{R}^{n_c \times s}$, and $\bar b^c \in \mathbb{R}^{n_c}$. Moreover, in the absence of the equality constraint (i.e., $\bar A^c=0$ and $\bar b^c=0$), the set $\cal{C}=\big<G,c\big>$  reduces to a zonotope. 
\end{defn}} 
\textcolor{blue}{\begin{defn} \textbf{\big(Matrix Zonotope (MZ) and Constrained MZ (CMZ)\big)} \cite{ID3}
Given a generator matrix $G \in \mathbb{R}^{n \times ms}$, a center $C \in \mathbb{R}^{n \times m}$, a CMZ of dimension $(n,m)$  with $s$ generators is represented by 
\begin{align*} 
& \cal{K}=\big<G,C,\bar A^C,\bar B^C\big>=\Big\{X \in \mathbb{R}^{n \times m}:X=\sum\limits_{i=1}^{s} G_{:,i}\zeta_i +C, \\ \nonumber & \quad \quad \quad \sum\limits_{i=1}^{s} \bar A^C_{:,i} \,  \zeta_i={\bar B^C}, \,\, \big\Vert \zeta  \big\Vert \leq 1  \Big\},
\end{align*}
where $G=[G_{:,1},...,G_{:,s}], \bar A^C=[\bar A^C_{:,1},...,\bar A^C_{:,s}]$ with $G_{:,i} \in \mathbb{R}^{n \times m}, \, \bar A^C_{:,i} \in \mathbb{R}^{n_c \times m_c}$ for $i=1,...,s$, ${\bar B^C} \in \mathbb{R}^{n_c \times m_c}$, and $\zeta=[\zeta_1,...,\zeta_s]^\top$. Moreover, in the absence of the equality constraint (i.e., ${\bar A^C}=0$ and ${\bar B^C}=0$), the set $\cal{K}=\big<G,C\big>$ reduces to an MZ.
\end{defn}}

\section{Operations on Constrained Matrix Zonotopes}
\textcolor{blue}{This section introduces intersection and transformation operations on CMZs. The intersection extends \cite{setop} to CMZs and requires new theoretical developments.} 

\begin{lem} \label{intmz}
    Consider two CMZs $\cal{M}^v=\big<G^v,C^v,A^v,B^v \big>$, $G^v \in \mathbb{R}^{n \times p s_v}$, $C^v \in \mathbb{R}^{n \times p}$, $A^v \in \mathbb{R}^{k_{v}  \times q_{v} s_v}$, and $B^v \in \mathbb{R}^{k_{v}  \times q_{v} }$, $v=1,2$.   Then, 
    \begin{align} \label{intmz1}
     &\cal{M}^1  \!\bigcap\! \cal{M}^2\! \!=\! \!\bigg<\! \![ G^1 \,\,\,\!  \mathbf{0}_{\scriptscriptstyle{n \times ps_2}}], \!C^1 \!, \!\!\begin{bmatrix}
               \! \! \!  \!\! \! \!\!\bar A^1 &\! \! \! \! \! \!\mathbf{0}_{\scriptscriptstyle{k_1 \times bs_2}} \! \! \\
            \mathbf{0}_{\scriptscriptstyle{k_2 \times bs_1}} &\!\!\! \bar A^2  \! \! \! \! \! \! \! \! \!\\
              \! \! \! \! \!\!\!\bar G^1 & \!\!\! \! \! - \bar G^2\!\! \! \! \! \! \end{bmatrix} \!, \! \!\begin{bmatrix} 
            \!\bar  B^1 \! \\
            \!\bar  B^2 \! \\
             \!\bar C^2- \bar C^1 \!
        \!\!\end{bmatrix}\! \!\bigg>,
    \end{align} 
    where $\bar A^v=[\bar A^v_{:,1} \quad ... \quad \bar A^v_{:,s_v}]$, $\bar G^v=[\bar G^v_{:,1} \quad ... \quad \bar G^v_{:,s_v}]$ with
    \begin{align} \label{BarA}
    &\bar  A^v_{:,i}=[A^v_{:,i} \,\,\, \mathbf{0}_{k_v \times (b-q_v)}], \bar  B^v=[B^v \,\,\, \mathbf{0}_{k_v \times (b-q_v)}] \in \mathbb{R}^{k_v \times b},   \nonumber \\ 
    &    \,\,
    \bar  G^v_{:,i}=[G^v_{:,i} \,\,\, \mathbf{0}_{n \times (b-p)}], \bar  C^v=[C^v \,\,\, \mathbf{0}_{n \times (b-p)}] \in \mathbb{R}^{n \times b},    \nonumber \\ & i=1,...,s_v,\,\, v=1,2,  
    \end{align}
    and $b=\max(p,q_1,q_2)$.
\end{lem}
\noindent \textit{Proof.} Let the right-hand side of \eqref{intmz1} be $\cal{M}_{int}$ and define 
\begin{align} \label{XZON}
   & B_{\infty}(A^v,B^v)= \bigg \{\zeta^v \,\, \Big|  \,\, \|\zeta^v\| \le 1, \sum_{i=1}^{s_v} A_{:,i}^v \, \zeta^v_i =B^v \bigg \} \nonumber \\ & = \bigg \{\zeta^v \,\, \Big|  \,\, \|\zeta^v\| \le 1,  A^v (\zeta^v \otimes I_{q_v}) = B^v \bigg \} \nonumber \\ & =\bigg \{ \zeta^v \,\, \Big|  \,\, \|\zeta^v\| \le 1,  \bar A^v ( \zeta^v \otimes I_{b}) =\bar B^v \bigg \}=B_{\infty}(\bar A^v,\bar B^v).
\end{align}

If $X \in \cal{M}^1 \bigcap \cal{M}^2$, since $X \in \cal{M}^1$, there exists a $\zeta^1 \in B_{\infty}(A^1, B^1)$ such that 
\begin{align} \label{X11}
  &  X=\sum_{i=1}^{s_1}  G_{:,i}^1 \, \zeta_i^1+ C^1=  G^1 (\zeta^1 \otimes I_{p})+ C^1.
\end{align}

On the other hand, since $X \in \cal{M}^2$, one has 
\begin{align*} 
   X= G^1 (\zeta^1 \otimes I_{p})+ C^1= G^2 (\zeta^2 \otimes I_{p})+ C^2, 
\end{align*}
for some $\zeta^2 \in B_{\infty}( A^2, B^2)$. Based on \eqref{BarA}, it yields 
\begin{align} \label{intXb}
    \bar G^1 (\zeta^1 \otimes I_{b})+ \bar C^1= \bar G^2 (\zeta^2 \otimes I_{b})+ \bar C^2. 
\end{align}

Let $\zeta=[{\zeta^1}^\top \quad {\zeta^2}^\top]^\top $.  Then, $\zeta \in B_{\infty}$, and from \eqref{X11}, $X=[ G^1 \quad \mathbf{0}_{n \times ps_2}] (\zeta \otimes I_{p})+C^1$. Moreover, $\zeta$ must satisfy $[\bar A^1  \quad \mathbf{0}_{k_1 \times bs_2}] (\zeta  \otimes I_{b}) =\bar B^1$, $[\mathbf{0}_{k_2 \times bs_1}  \quad \bar A^2] (\zeta  \otimes I_{b}) =\bar B^2$ (based on \eqref{XZON}), and $[\bar G^1 \quad -\bar G^2] (\zeta \times I_b)=\bar C^2-\bar C^1$ (based on \eqref{intXb}), which can compactly be written as 
\begin{align} \label{newAB}
    \begin{bmatrix}
            \bar A^1 & \mathbf{0}_{k_1 \times bs_2} \\
            \mathbf{0}_{k_2 \times bs_1} & \bar A^2 \\
            \bar G^1 & - \bar G^2
        \end{bmatrix}  (\zeta \otimes I_b) =\begin{bmatrix}
            \bar B^1 \\
           \bar  B^2 \\
            \bar C^2- \bar C^1
        \end{bmatrix}.
\end{align}

 Since $\bar A^v_{:,i}$, $\bar G^v_{:,i}$, $\bar B^v$, and $\bar C^2- \bar C^1$ (for $i=1,..,s_v$, $v = 1,2$) all have $b$ columns, \eqref{newAB} is valid. Therefore, $X \in \cal{M}_{int}$. Conversely, if $X \in \cal{M}_{int}$, there exists $\zeta \in B_{\infty}$ satisfying \eqref{newAB} such that $X=[ G^1 \quad \mathbf{0}_{n \times ps_2}] (\zeta \otimes I_{p})+ C^1$. Letting $\zeta=[{\zeta^1}^\top \quad {\zeta^2}^\top]^\top $, it follows that $X= G^1 (\zeta^1 \otimes I_{p})+ C^1$
with $\zeta^1 \in B_{\infty}(\bar A^1,\bar B^1)=B_{\infty}( A^1, B^1)$. Therefore, $X \in \cal{M}^1$. Besides, the equality constraint from $\cal{M}_{int}$ gives $\bar G^1 (\zeta^1 \otimes I_{b})+\bar C^1=\bar G^2 (\zeta^2 \otimes I_{b})+\bar C^2$ with $\zeta^2 \in B_{\infty}( \bar A^2, \bar B^2)$, implying $ G^1 (\zeta^1 \otimes I_{p})+ C^1= G^2 (\zeta^2 \otimes I_{p})+C^2$ with $\zeta^2 \in B_{\infty}( A^2, B^2)$. Therefore, if $X \in \cal{M}^1$, then $X \in \cal{M}^2$. This completes the proof. \hfill   $\blacksquare$ 

\begin{lem} \label{trans2}
    Let $\cal{K}=\big <G,C,\bar A^C,\bar B^C \big>$ be a CMZ of dimension $(n,m)$. Then,  $\cal{K}\circ N =\{XN: X \in \cal{K}\}$ for some vector $N \in \mathbb{R}^{m}$ (matrix $N \in \mathbb{R}^{m \times q}$  ) is a constrained zonotope (a 
    CMZ), defined by $\cal{K}_N=\big <G  \circ N,CN,\text{Vec}(\bar A^C),\text{Vec}(\bar B^C) \big>$ \big($\cal{K}_N=\big <G \circ N,CN,\bar A^C,\bar B^C \big>$\big). 
\end{lem}
\noindent \textit{Proof:} The proof follows the approach in \cite{Czen}. \hfill   $\blacksquare$


\section{Problem Formulation}
Consider the discrete-time system with dynamics 
\begin{equation}\label{system} 
x(t+1) = {A}^{*}x(t) + B^{*} u(t) + w(t),
\end{equation}
where $x(t) \in \cal{S}_s \subset \mathbb{R}^n$ is the system's state, $u(t) \in \mathbb{R}^m$ is the control input, and $w(t) \in  \cal{Z}_w \subset \mathbb{R}^n$ is the additive disturbance. 


\begin{assumption} \label{PK}
 $\theta^{*}=[A^{*} \quad B^{*}]$ is uncertain and $\theta^* \in \cal{M}_{prior}=\big <G^{\theta},C^{\theta},A^{\theta},B^{\theta} \big>$ for some $C^{\theta} \in \mathbb{R}^{n \times (n+m)}$ and $G^{\theta} \in \mathbb{R}^{n \times (n+m)s_{\theta}}$, $A^{\theta} \in \mathbb{R}^{n_{\theta} \times m_\theta s_{\theta}}$ and $B^{\theta} \in \mathbb{R}^{n_{\theta} \times m_\theta}$. 

\end{assumption}
\textcolor{blue}{\begin{assumption} \label{contr}
 The pair $(A^*,B^*) \in \cal{M}_{prior}$ is stabilizable. 
\end{assumption}}
\begin{assumption} \label{dist}
The disturbance set $\cal{Z}_w$ is a zonotope. That is, $\cal{Z}_w=\big<G^h,c^h\big>$ for some $G^h \in \mathbb{R}^{n \times s_{w}}$, and $c^h \in  \mathbb{R}^{n}$.\end{assumption} 
\vspace{-8pt}
\textcolor{blue}{\begin{assumption} \label{safeset}
  The system's safe set is given by a polytope
$
\cal{S}_s=\cal{P} (H,h) = \big \{ x \in {\mathbb{R}^n}:H x  \le h , \,\, H \in \mathbb{R}^{q \times n}, \,\, h \in \mathbb{R}^q\big\}.
$
\end{assumption}} 
To collect data for learning, a sequence of control inputs is applied to the system \eqref{system} as
\begin{align} \label{data-u}
U^0 := \begin{bmatrix} u(0) & u(1) & \cdots & u(T-1) \end{bmatrix} \in \mathbb{R}^{m \times T}.
\end{align}


The collected state samples are then organized as  
\begin{align} \label{data-x}
X^0 &:= \begin{bmatrix} x(0) & x(1) & \cdots & x(T-1) \end{bmatrix} \in \mathbb{R}^{n \times T}, \\
X^1 &:= \begin{bmatrix} x(1) & x(2) & \cdots & x(T) \end{bmatrix} \in \mathbb{R}^{n \times T}.  \label{data-z}
\end{align}

The sequence of unknown disturbances is 
\begin{equation*}
W^0 :=  \begin{bmatrix} w(0) & w(1) & \ldots & w(T-1) \end{bmatrix} \in \mathbb{R}^{n \times T}.
\end{equation*}

Based on Assumption \ref{dist}, the MZ formed by $T$-concatenation of $\cal{Z}_w$ is \cite{ID3} 
\begin{align} \label{T-dis}
    \cal{M}_{\cal{Z}_w^T}=\big <G^w,C^w\big>,
\end{align}
where $G^w = [\,G^w_{:,1},\,\ldots,\,G^w_{:,T s_w}\,]$ with each $G^w_{:,i} \in \mathbb{R}^{n\times T}$, and $C^w \in \mathbb{R}^{n \times T}$. One has $W^0 \in \cal{M}_{\cal{Z}_w^T}$.

We also define
$
  D^0:= \begin{bmatrix}
X^0 \\ U^0
\end{bmatrix} \in \mathbb{R}^{(n+m) \times T}.
$
\begin{assumption}\label{assumption_5}
The data matrix $D^0$ has full row rank.
\end{assumption} 
\noindent \textbf{Problem 1: Data-based Safe Control Design} 
Consider the collected data \eqref{data-u}-\eqref{data-z} from the system \eqref{system} under Assumptions 1-5. Learn a controller in the form of 
$u(t)=K x(t)$,
to ensure that the safe set described in Assumption 4 is a robust invariant set (RIS) (i.e., if $x(0) \in \cal{P}$, then $x(t) \in \cal{P} \,\,\, \forall t \ge 0, \,\,\,  \forall w(t) \in \cal{Z}_w $ \cite{SetB}).




\section{Closed-loop System Representation Using Data and Prior Knowledge}
In this section, we present a novel data-based closed-loop representation that integrates data and prior knowledge. 

We denote the set of all such open-loop models consistent with both data and prior knowledge by
\begin{align*} 
    \Sigma_{X,U^0}=\Big \{\theta \in \cal{M}_{prior}: X^1=\theta D^0+W^0, \,\, W^0 \in  \cal{M}_{\cal{Z}_w^T} \Big  \},
\end{align*}

\noindent where $\theta=[A \quad B]$. For a gain $K$, define the set of closed-loop systems consistent with data and prior knowledge as
\begin{align} \label{cl-set}
   \Sigma^K_{X,U^0}= \Big \{A^K: A^K=A+BK, \,\,  [A \quad B] \in \Sigma_{X,U^0} \Big \}.
\end{align}

Inspired by \cite{Data2}, we parameterize the control gain $K$ by a decision variable $V^K \in \mathbb{R}^{T \times n}$ that must satisfy 
\begin{align} \label{G}
 \begin{bmatrix}
    I_n \\ K
\end{bmatrix}= \begin{bmatrix}
     X^0  \\   U^0 
\end{bmatrix}   V^K  = D^0   V^K.
\end{align}

Unlike \cite{Data2}, we present a CMZ-based representation of the closed-loop system by integrating prior knowledge with data.

\begin{thm} \label{clrep}
Consider the system \eqref{system} under Assumptions 1-3 and 5. Let the input-state collected data be given by \eqref{data-u}-\eqref{data-z}. Then, under parametrization \eqref{G}, the set \eqref{cl-set} is exactly represented by the following CMZ of dimension $(n,n)$
 \begin{align}\label{clset}
& \cal{M}_{cl}\!=\!\Big<\!\Big[- G^w \circ V^K \!\quad\! \mathbf{0}_{n \times ns_{\theta}} \Big], \Big(X^1-C^w\Big)V^K,   A^C,B^C \!\Big>,
\end{align}
with  $s_c=s_{\theta}+Ts_w$ generators and
\begin{align}  \label{AcBc}
  A^C=\begin{bmatrix}
        \bar A^w  & \mathbf{0} \\ 
        \mathbf{0} & \bar A^{\theta} \\
      \bar G^w & -\bar G^\theta
    \end{bmatrix}, \quad B^C=\begin{bmatrix}
        \bar B^w  \\ 
        \bar B^{\theta} \\
       \bar C^{\theta}-\bar C^w
    \end{bmatrix} ,  
\end{align}
where the parameters of $A^C$ and $B^C$ are defined in the table below with $p = n+m$, and
\begin{subequations}\label{Aw}
\begin{align} 
 &   A^{w}=G^{w}\circ (D^0)^{\bot} \in \mathbb{R}^{n \times (T-p) Ts_w}, \\ & B^{w}=\Big[ X^1- C^w  \Big]    (D^0)^{\bot} \in \mathbb{R}^{n \times (T-p)}.
\end{align}
\vspace{-6pt}
\begin{table}[h]
    \centering
    \renewcommand{\arraystretch}{1.2}
    \setlength{\tabcolsep}{4pt} 
    \begin{tabular}{|c|c|c|}
        \hline
        Matrix & $m_\theta < T$ & $m_\theta > T$ \\ \hline
        $\bar{A}^{\theta}$ & $[A^{\theta} \quad \mathbf{0}_{n \times (T - m_\theta)s_\theta}]$ & $A^{\theta}$ \\ \hline
        $\bar{B}^{\theta}$ & $[B^{\theta} \quad \mathbf{0}_{n \times (T - m_\theta)}]$ & $B^{\theta}$ \\\hline 
        $\bar{A}^w$ & $[A^w \quad \mathbf{0}_{n \times pTs_w}]$ & $[A^w \quad \mathbf{0}_{n \times (m_\theta - (T-p))Ts_w}]$ \\ \hline
        $\bar{B}^w$ & $[B^w \quad \mathbf{0}_{n \times p}]$ & $[B^w \quad \mathbf{0}_{n \times (m_\theta - (T-p))}]$ \\ \hline
        $\bar{G}^w$ & $ G^w$ & $[ G^w \quad \mathbf{0}_{n \times (m_\theta-T) Ts_w} ]$ \\ \hline
        $\bar{C}^w$ & $C^w $ & $[ C^w \quad \mathbf{0}_{n \times (m_\theta-T)}]$ \\ \hline
        $\bar{G}^{\theta}$ & $-G^{\theta} \circ D^0 $ & $[-G^{\theta} \circ D^0 \quad \mathbf{0}_{n \times (m_\theta-T) s_\theta} ]$ \\ \hline
        $\bar{C}^{\theta}$ & $ {X}^1-C^{\theta} D^0 $ & $[ {X}^1-C^{\theta} D^0 \quad \mathbf{0}_{n \times (m_\theta-T)}]$ \\ \hline
    \end{tabular}
    \caption{ \small{The parameters of $A^C$ and $B^C$}}
    \label{Aug}
\end{table}
\end{subequations}

Moreover, under Assumption \ref{assumption_5}, the set $\cal{M}_{cl}$ is non-empty and compact.
\end{thm}
\noindent \textit{Proof: }
Using the data \eqref{data-u}-\eqref{data-z} in \eqref{system}, we have
\begin{equation}\label{system-clN1} 
X^1 = {A}^*X^0 + B^*U^0 + W^0.
\end{equation}

Using  $K= U^0 V^K$ and $X^0 V^K=I_n$
from \eqref{G} in \eqref{system-clN1}, the data-based closed-loop dynamics simplifies to 
\begin{align}\label{system-cl2} 
A^*+B^*K=({X}^1-{W}^0) V^K.
\end{align}

From \eqref{T-dis}, the closed-loop system set $A^K = (X^1 - W^0)V^K$ forms a MZ. To compute the exact set $\Sigma^K_{X,U^0}$, we refine it by eliminating disturbances incompatible with data and prior knowledge. Using \eqref{G}, we get
\begin{align} \label{sys-cl3}
 A^*+B^*K= \begin{bmatrix}
    A^* & B^*
 \end{bmatrix} \begin{bmatrix}
    I_n \\ K
 \end{bmatrix}=  \begin{bmatrix}
    A^* & B^*
 \end{bmatrix} D^0 \, V^K.
\end{align}

Comparing \eqref{system-cl2} and \eqref{sys-cl3}, the data consistency condition $[A \quad B]=\theta \in \Sigma_{X,U^0}$ implies that 
\begin{align} \label{con11}
 \theta D^0 = X^1- W^0,
\end{align}
which holds for some $ W^0 \in \cal{M}_{\cal{Z}_{{w}}^{T}}$ and some $\theta=[A \quad B] \in \cal{M}_{prior}$.
Using \eqref{T-dis} and based on Fredholm alternative \cite{ID5,ID3}, there exists a solution to \eqref{con11} if and only if
\begin{align*}
\Big [ X^1- C^w  \Big]    (D^0)^{\bot}=\sum\limits_{i=1}^{Ts_w} G^{w}_{:,i} \, (D^0)^{\bot}  \zeta_i,
\end{align*}
with $\big\Vert \zeta  \big\Vert \leq 1$, $\zeta \in \mathbb{R}^{Ts_{w}}$. Imposing this equality reduced the MZ \eqref{T-dis} for the disturbance sequence to the CMZ
$\cal{M}_{W}=\Big< G^w, C^w  ,A^w,B^w \Big>$,
with $A^w$ and $B^w$ defined in \eqref{Aw}. 

We now exclude the disturbances for which the resulting solutions $\theta$ are not consistent with the prior knowledge. Based on \eqref{con11} and Assumption~\ref{PK}, the disturbance set that aligns with prior knowledge is given by
\begin{align*} 
&  W^0 \!=\!  X^1-\theta  D^0 \in \!\Big<\! -G^{\theta} \circ D^0, X^1-C^{\theta}D^0 ,A^{\theta},B^{\theta} \!\Big>  \!=\! \cal{M}_{d},
\end{align*}
where Lemma \ref{trans2} is used to find $\theta D^0$. Moreover, $G^{\theta}\circ D^0 \in \mathbb{R}^{n \times Ts_{\theta}}$ and $ {X}^1-C^{\theta} D^0 \in \mathbb{R}^{n \times T}$. 
Hence, the disturbance set consistent with prior knowledge and explainable by data is given by the intersection ${W}^0 \in \cal{M}_W \bigcap \cal{M}_{d}=\cal{M}_{dp}$. 
The number of columns in $B^w$, $B^{\theta}$, and the generators $G^w_{:,i}$ and $[-G^{\theta} \circ D^0]_{:,i}$ are $q_1 = T - p$ (with $p = m + n$), $q_2 = m_{\theta}$, and $q_3 = T$, respectively. To apply Lemma~\ref{intmz} for intersecting $\cal{M}_W$ and $\cal{M}_d$, we ensure dimensional consistency. Since $q_1 < q_3$, if $q_2 > q_3$, then $G^{w}_{:,i}, \, i=1,...,Ts_{w}$, $C^w$, $[-G^{\theta} \circ D^0]_{:,i}, \, i=1,...,s_{\theta}$, and $[ {X}^1-C^{\theta} D^0]$ are padded with $\mathbf{0}_{n \times (m_\theta - T)}$. Besides, $A^{w}_{:,i}, \, i=1,...,Ts_{w}$ and $B^{w}$ are padded with $\mathbf{0}_{n \times (m_\theta-(T - p))}$. On the other hand, if $q_2 < q_3$, then $A^{w}_{:,i}, \, i=1,...,Ts_w$ and $B^w$ are padded with $\mathbf{0}_{n \times p}$, and $A^{\theta}_{:,i}, \, i=1,...,s_{\theta}$ and $B^{\theta}$ are padded $\mathbf{0}_{n \times (T-m_\theta)}$ to ensure consistency. Then, using Lemma \ref{intmz} and Table~\ref{Aug}, it yields
\begin{align} \label{dp}
    \cal{M}_{dp}= &\Bigg< \Big[G^w  \!\quad\! \mathbf{0}_{n \times Ts_{\theta}} \Big], C^w , \begin{bmatrix}
        \bar A^w  & \mathbf{0}  \\ 
        \mathbf{0} & \bar A^{\theta} \\
       \bar G^w & -\bar G^\theta
    \end{bmatrix},\begin{bmatrix}
        \bar B^w  \\ 
       \bar B^{\theta} \\
       \bar C^\theta - \bar C^w 
    \end{bmatrix} \Bigg>.
\end{align}

Therefore, $\Sigma^K_{X,U^0} \subseteq \cal{A}_G$, where
\begin{align} \label{setcl2}
 &  \cal{A}_G= \Big \{A^K: A^K=( X^1- W^0) V^K, \,\,  W^0 \in \cal{M}_{dp} \Big \}.
\end{align}

Applying Lemma \ref{trans2} and substituting \eqref{dp} into \eqref{setcl2} leads to \eqref{clset}. Moreover, by Assumption \ref{assumption_5}, there exists $V^K$ for every $K$, ensuring that the set is non-empty. \hfill   $\blacksquare$ 


\section{Unifying Direct and Indirect Learning}
\textcolor{blue}{To unify direct learning (i.e., closed-loop learning) with indirect learning (i.e., open-loop learning), in this section, we develop a novel zonotope-based system identifier, which adaptively replaces $\cal{M}_{prior}$ in the closed-loop learning process.}





Set-based identification using matrix ellipsoidal model sets ~\cite{QMI,Trans} often results in conservatism and leads to semi-definite programs in control design. To resolve these issues, we introduce a CMZ-based identification. Zonotope-based identification is limited to scalar systems~\cite{ID2,ID2z}.

The system identifier relies on adaptively refining the set of prior knowledge after a batch of data is collected. Let the batch of data collected be
\begin{align} \label{data-us}
U^{s0} := \begin{bmatrix} u(0) & u(1) & \cdots & u(T^s-1) \end{bmatrix} \in \mathbb{R}^{m \times T^s},
\end{align} 
\begin{align} \label{data-xs}
X^{s0} &:= \begin{bmatrix} x(0) & x(1) & \cdots & x(T^s-1) \end{bmatrix} \in \mathbb{R}^{n \times T^s}, \\
X^{s1} &:= \begin{bmatrix} x(1) & x(2) & \cdots & x(T^s) \end{bmatrix} \in \mathbb{R}^{n \times T^s}, \label{data-zs}
\end{align}
\begin{equation*}
W^{s0} :=  \begin{bmatrix} w(0) & w(1) & \ldots & w(T^s-1) \end{bmatrix} \in \mathbb{R}^{n \times T^s}.
\end{equation*}
where $T^s$ is the number of samples.


Using the collected data and the system \eqref{system}, we have
\begin{equation*}
X^{s1} = {A}^*X^{s0} + B^*U^{s0} + W^{s0}.
\end{equation*}

Using $W^{s0}\!=\!X^{s1}\!-\!{A}^*X^{s0}\!-\!B^*U^{s0}$, the matrix interval of the MZ $W^{s0}$ defined as in \eqref{T-dis}, $\cal{M}_{\cal{Z}^{T^s}_{w}}\!=\!\big <\!G^{sw},C^{sw}\!\big>$, becomes $\cal{I}_{W^{s0}}\!=\!\big[C^{sw}\!-\!\Delta G^{sw} \!\quad\! C^{sw}\!+\!\Delta G^{sw} \big]$, with $\Delta G^{sw} = \textstyle\sum_{i=1}^{T^ss_w} |G^{sw}_{:,i}|$ \cite{Althoff}. Then, the feasible open-loop system model set (FOLMS) of $(A,B)$ in \eqref{system} that can explain the data is
\begin{align}\label{IDset} 
\cal{L}_d=\bigg \{\theta: \bigg|\theta D^{s0}- X^{s1}+C^{sw} \bigg| \le \Delta G^{sw}  \bigg \}.
\end{align}\textcolor{blue}{\begin{rem}Note that $ D^{s0}:= \begin{bmatrix}
X^{s0} \\ U^{s0}
\end{bmatrix}$ is not required to be full row rank, since the information set $\cal{L}_d$ need not be compact. This is because the FOLMS in \eqref{IDset} serves to refine the prior knowledge set, which initializes the FOLMS. As new data are collected, a posterior FOLMS is obtained by intersecting \eqref{IDset} with the prior FOLMS. The following lemmas are developed to enable intersecting the prior knowledge or prior FOLMS with the information set in \eqref{IDset} to compute the posterior FOLMS.
\end{rem}} 

\begin{lem}
The CMZ $\cal{M}=\big<G,C,\bar A^C,\bar B^C\big>$ of dimension $(n,m)$ with $\bar A^C\in \mathbb{R}^{n_c \times m_cs}$ and $\bar B^C\in \mathbb{R}^{n_c \times m_c}$ intersects $\mathcal H= \big \{ \bar \theta \in \mathbb{R}^{n \times m}: \bar \theta X = F \big \}$ with $X \in \mathbb{R}^{m \times m}$ and $F \in \mathbb{R}^{n \times m}$ if and only if 
there exists a $\zeta$ that satisfies
\begin{subequations} \label{search1}
\begin{align} 
\label{search1a}
& \left(G(\zeta \otimes I_m) + C \right) X = F, \\
\label{search1b}
& \|\zeta\| \leq 1, \\
& \bar A^C (\zeta \otimes I_{m_c}) = \bar B^C.
\end{align}
\end{subequations}
\end{lem} 
\noindent \textit{Proof:} We first show the necessity. Assume $\cal{M} \cap \mathcal H \neq \emptyset$, i.e., there exists a $\bar \theta \in \cal{M}$ satisfying $\bar \theta X = F$. Therefore, using the definition of CMZ, there exists $\zeta$ such that
\[
\left(G(\zeta \otimes I_m) + C \right) X = F, \!\!\!\quad \|\zeta\| \leq 1, \!\!\!\quad \bar A^C (\zeta \otimes I_{m_c}) = \bar B^C.
\]

We now show the sufficiency. Assume that \eqref{search1} has a solution $\bar \zeta$.  
Then, we show that $(G(\bar \zeta \otimes I_m) + C) \in \cal{M} \cap \mathcal H \neq \emptyset$. The inclusion $(G(\bar \zeta \otimes I_m) + C) \in \cal{M}$ is immediate from \eqref{search1}, since
$\bar A^C (\bar \zeta \otimes I_{m_c}) = \bar B^C$ for $\|\bar \zeta\| \leq 1$.
Besides, $(G(\bar \zeta \otimes I_m) + C) \in \mathcal{H}$, since  
\(
(G(\bar \zeta \otimes I_m) + C) X = F.
\)
This completes the proof.
\hfill   $\blacksquare$ 
\begin{lem} \label{intLemma}
    Consider the CMZ $\cal{M}=\big<G,C,\hat A^C,\hat B^C\big>$ of dimension $(n,m)$ with $\hat A^C\in \mathbb{R}^{n_c \times ms}$ and $\hat B^C\in \mathbb{R}^{n_c \times m}$, and the set $\hat {\mathcal H}= \big \{ \bar \theta \in \mathbb{R}^{n \times m}: \bar\theta X \le F \big \}$ with $X \in \mathbb{R}^{m \times m}$ and $F \in \mathbb{R}^{n \times m}$. Let \eqref{search1} be satisfied. Then, the intersection $\cal{M}_{int}=\cal{M} \bigcap \hat {\mathcal H}$ is a CMZ given by
\begin{align}\label{Mint}
    \cal{M}_{int}\!=\!\Big<\![G \quad\! \mathbf{0}_{\scriptscriptstyle{n\times m}}],\!C,\!\begin{bmatrix}
        \!\hat A^C  &\!\! \mathbf{0}_{\scriptscriptstyle{n_c \times   m}}\!\\
        \!G \circ X &\!\! \frac{L^M}{2}\!
    \end{bmatrix}\!,\!\begin{bmatrix} \hat B^C \\ F-CX-\frac{L^M}{2} \end{bmatrix}\!\Big>,
\end{align}
where
\begin{align}\label{LM}
    L^M=F-CX+\sum_{i=1}^s \big |G_{:,i} \, X \big|.
\end{align}
\end{lem}
\noindent \textit{Proof:} If $\bar\theta \in \cal{M}_{int}$, we show that $ \bar\theta \in \cal{M} \cap \hat{\mathcal H} $. From $ \cal{M}_{int} $ in \eqref{Mint}, there exists a $ \zeta \in \mathbb{R}^s $ and a $ \zeta_{s+1} \in \mathbb{R} $ such that
\[
\bar\theta = G \zeta + \mathbf{0}_{n \times m} \, \zeta_{s+1} + C, \quad \|\zeta\| \leq 1, \quad |\zeta_{s+1}| \leq 1,
\]
and the following conditions hold 
\begin{subequations} \label{condLM}
\begin{align}
\label{1condLM}
&\sum_{i=1}^s G_{:,i} \, X \, \zeta_i + \frac{L^M}{2} \zeta_{s+1} = F - CX - \frac{L^M}{2},\\
\label{2condLM1}
&\sum_{i=1}^s \hat A^{C}_{:,i} \, \zeta_i = \hat B^C.
\end{align}
\end{subequations}

If $\cal{M} \cap \hat{\mathcal H} \neq \emptyset $, then from \eqref{search1a}, $|F-CX|\leq \sum_{i=1}^{s}|G_{:,i}X|$ due to \eqref{search1b}. Therefore, $L^M \geq \mathbf{0}_{n \times m}$ in \eqref{LM}. If $L^M = \mathbf{0}_{n \times m}$, \eqref{1condLM} simplifies to
$
\sum_{i=1}^s G_{:,i} \, X \, \zeta_i = F - CX,
$
which can be rearranged as $
(C + \sum_{i=1}^s G_{:,i} \, \zeta_i) X = F$. Therefore, $ \bar\theta \in \cal{M}_{int} \subset \cal{M} $ and $ \bar\theta \in \mathcal H \subset \hat{\mathcal H} $. If $L^M > \mathbf{0}_{n \times m}$, then we solve \eqref{condLM} for $ L^M\zeta_{s+1} $ as
\begin{equation} \label{LMzetas}
L^M\zeta_{s+1} = 2(F - CX - \frac{L^M}{2} - \sum_{i=1}^s G_{:,i} \, X \, \zeta_i).
\end{equation}

By multiplying both sides of \eqref{LMzetas} by $(L^M)^\top$ and taking the trace on both sides, $\zeta_{s+1}$ can be drived as
\begin{equation}\label{zetas}
\zeta_{s+1} \!=\! \frac{2 \, \text{trace}\left( (L^M)^\top\! \left( F - CX - \frac{L^M}{2} - \sum_{i=1}^s G_{:,i} \, X \, \zeta_i \right) \right)}{\text{trace}((L^M)^\top L^M)}.
\end{equation}

Combining \eqref{LMzetas} with $ -1 \leq \zeta_{s+1} $, we obtain
\[
-L^M \leq L^M\zeta_{s+1} = 2(F - CX - \frac{L^M}{2} - \sum_{i=1}^s G_{:,i} \, X \, \zeta_i),
\]
which leads to
\(
-\frac{L^M}{2} \leq F - CX - \frac{L^M}{2} - \sum_{i=1}^s G_{:,i} \, X \, \zeta_i,
\)
and simplifies to
\(
(C + \sum_{i=1}^s G_{:,i} \, \zeta_i) X \leq F.
\)
Thus, $ \bar\theta \in \cal{M} $ and $ \bar\theta \in \hat{\mathcal H} $. 
Let \( \bar\theta \in \cal{M} \cap \hat{\mathcal H} \). To complete the proof, we show that $ \bar\theta \in \cal{M}_{int} $. Since \( \bar\theta \in \cal{M} \cap \hat{\mathcal H} \), there exists \( \zeta \in \mathbb{R}^s \) such that
\begin{align}
\label{condLM2}
    \bar\theta = G \zeta + C, \,\,  \sum_{i=1}^s \hat A^{C}_{:,i} \, \zeta_i= \hat B^C, \,\, \|\zeta\| \leq 1, \,\,
    \bar\theta X \leq F.
\end{align}

To show \( \bar\theta \in \cal{M}_{\text{int}} \), it suffices to find a \( \zeta_{s+1} \in \mathbb{R} \) such that
\[
\bar\theta = G \zeta + \mathbf{0}_{n \times m} \, \zeta_{s+1} + C, \quad |\zeta_{s+1}| \leq 1,
\]
and \eqref{condLM} holds.
If \( L^M = \mathbf{0}_{n \times m} \), then by \eqref{LM}, \( F - CX = -\sum_{i=1}^s |G_{:,i} \, X| \). Using \eqref{condLM2}, it follows that \( \sum_{i=1}^s G_{:,i} \, X \, \zeta_i \leq F - CX \), and thus
\(
\sum_{i=1}^s G_{:,i} \, X \, \zeta_i \leq -\sum_{i=1}^s |G_{:,i} \, X|.
\)
For \( \|\zeta\| \leq 1 \), one has
\[
-\sum_{i=1}^s |G_{:,i} \, X| \le \sum_{i=1}^s G_{:,i} \, X \, \zeta_i \leq \sum_{i=1}^s |G_{:,i} \, X|.
\]

Consequently, 
$
 \sum_{i=1}^s G_{:,i} \, X \, \zeta_i = -\sum_{i=1}^s |G_{:,i} \, X| = F - CX,
$
which satisfies \eqref{1condLM} for \( L^M = \mathbf{0}_{n \times m} \), regardless of \( \zeta_{s+1} \). 
Hence, \( \zeta_{s+1} \) can be freely chosen as long as \( |\zeta_{s+1}| \leq 1 \). If \( L^M > \mathbf{0}_{n \times m} \), we choose \( \zeta_{s+1} \) as in \eqref{zetas}, which ensures \eqref{LMzetas} and \eqref{1condLM}. To show \( |\zeta_{s+1}| \leq 1 \), consider \( \bar\theta \) from \eqref{condLM2}. Since \( F - \bar\theta X \geq \mathbf{0}_{n \times m} \), it follows that \( \zeta_{s+1} \) satisfies
\[
L^M\zeta_{s+1} = 2 \left( F - CX - \sum_{i=1}^s G_{:,i} \, X \, \zeta_i - \frac{L^M}{2} \right) \geq -L^M,
\]
which leads to $\zeta_{s+1}  \geq -1$. Using the fact that \( - \sum_{i=1}^s G_{:,i} \, X \, \zeta_i \leq \sum_{i=1}^s |G_{:,i} \, X| \) along with \eqref{condLM2}, we obtain
\[
L^M\zeta_{s+1} \leq 2\left( F - CX + \sum_{i=1}^s |G_{:,i} \, X| - \frac{L^M}{2} \right).
\]

Using the definition of \( L^M \) from \eqref{LM} yields
\[
L^M\zeta_{s+1} \leq 2\left( L^M - \frac{L^M}{2} \right) = L^M.
\]

Thus, $\zeta_{s+1}  \leq 1$. Therefore, for every \( \bar \theta \in \cal{M} \cap \hat{\mathcal H} \), it follows that \( \bar \theta \in \cal{M}_{int} \).\hfill   $\blacksquare$ 

We now define the set of systems consistent with the data used for open-loop learning and prior knowledge by 
\begin{align} \label{Ldp}
    \cal{L}_{dp}=\bigg \{\theta: \theta \in \cal{M}_{prior} \bigcap \cal{L}_d \bigg \},
\end{align}
where $\cal{M}_{prior}=\big<G^{\theta},C^{\theta},A^{\theta},B^{\theta}\big>$ based on Assumption \ref{PK}, and $\cal{L}_d$ is defined in \eqref{IDset}.
\begin{lem} \label{LdpLem}
    Let $2T^s>m_\theta$. The set 
    $\cal{L}_{dp}$  in \eqref{Ldp} is a CMZ given by
\begin{align*}
 &\cal{L}_{dp} =\\&\bigg<\!\big [G^{\theta} \quad \mathbf{0}_{n \times (n+m)}],C^{\theta},\begin{bmatrix}
\begin{bmatrix}A^\theta \!\quad\! \mathbf{0}_{n_\theta \times (2T^s-m_\theta)s_\theta}\end{bmatrix} & \mathbf{0}_{n_\theta \times 2T^s}\\G^{\theta} \circ \begin{bmatrix}
     D^{s0} &
    - D^{s0}
\end{bmatrix} & \frac{\bar L^M}{2}
\end{bmatrix}, \nonumber \\ &  
\begin{bmatrix} \begin{bmatrix}B^\theta\quad \mathbf{0}_{n_\theta \times (2T^s-m_\theta)}\end{bmatrix}\\ 
    [F^1 \!\quad\! F^2] \! - \! C^{\theta} [D^{s0} \!\!\quad\!\! -\! D^{s0}]\! -\! \frac{\bar L^M}{2}
\end{bmatrix}\!\bigg> \! = \! \bigg<\!\!{G}^{d\theta}\!,\!C^{d\theta}\!,\!{A}^{d\theta}\!,\!{B}^{d\theta}\!\!\bigg>,
\end{align*}
where ${A}^{d\theta} \in \mathbb{R}^{(n+n_\theta) \times {2T^s} (s_{\theta}+1) }$, ${B}^{d\theta} \in \mathbb{R}^{(n+n_\theta) \times {2T^s}}$, $F^1 = X^{s1}-C^{sw}+\Delta G^{sw}$, $F^2 = -X^{s1}+C^{sw}+\Delta G^{sw}$, and $\bar L^M = [F^1 \quad F^2]-C^{\theta} [D^{s0} \quad -D^{s0}]+\sum_{i=1}^{s_{\theta}} |G^{{\theta}}_{:,i} [D^{s0} \quad -D^{s0}]|$. 
\end{lem}
\noindent \textit{Proof:} The set in \eqref{IDset} consists of all \( \theta \) satisfying  
\( \theta D^{s0} \le X^{s1} - C^{sw} + \Delta G^{sw} \) and  
\( \theta (-D^{s0}) \le -X^{s1} + C^{sw} + \Delta G^{sw} \).
 The remainder follows from Lemma~\ref{intLemma}. \hfill   $\blacksquare$ 

\section{Robust Safe Control Design} 
\textcolor{blue}{
This section uses the learned closed-loop CMZ model to guarantee robust invariance of the safe set using the concept of $\lambda$-contractivity~\cite{SetB}. A set $\cal{P}$ is $\lambda$-contractive for system \eqref{system} if, for any $\lambda \in (0,1)$, \( x(t) \in \cal{P} \) implies \( x(t+1) \in \lambda \cal{P} \) for all \( t > 0 \) and all \( w \in \cal{Z}_w \).
}

\begin{thm}\label{th3}
Consider the system \eqref{system} under Assumptions 1-5. Then, Problem 1 is solved using the controller $u(t)= U^0 V^K x(t)$ if there exist $V^K$, $\rho$, r and $P$ satisfying 
\begin{subequations}
\label{LPf}
\begin{align} 
  \textcolor{blue}{\min\limits_{P,V^K,\rho,r}} \,\,\, & \textcolor{blue}{r}  \\
  {\rm{s}}{\rm{.t}}{\rm{.}} \quad &{P} h \le \lambda  h-H c^h-\rho \, l -y,   \\
&{P} H  =  H (X^1-C^w) V^K,  \\
& \big\Vert V^K \big\Vert  \le \rho,  \\
& X^0 V^K=I_n, \\
& P \ge 0,
\end{align} 
\end{subequations}
where $y=[y_1,...,y_q]^{\top}$ and $l=[l_1,...,l_q]^{\top}$, and
\begin{align} \label{yj}
& y_j  = \sum\limits_{i=1}^{s_w}  \Big|{H}_{j,:}  \,G^h_{:,i} \Big|, 
\end{align} 
\begin{subequations}
\label{lj}
\begin{align}
  l_j  = {M_x}\max\limits_{\beta} \,  & \big\|\sum\limits_{i=1}^{Ts_w}   \beta_i {H}_{j,:} \, G^{w}_{:,i}  \big\Vert     \,   \\ {\rm{s}}{\rm{.t}}{\rm{.}}\quad & \sum\limits_{i=1}^{s_c} {A}^{C}_{:,i} \, \beta_i={B}^{C}, \label{Eq1} \\  
  & \big\Vert \beta  \big\Vert \leq 1, 
\end{align}
\end{subequations}
with $M_x$ is the bound on $\|x(t)\|$, $s_c = s_{\theta}+Ts_w$, $A^C$ and $B^C$ defined in \eqref{AcBc}. \textcolor{blue}{When $\lambda$ is fixed, $r = \rho$; otherwise, $r = \lambda$ with $0 < \lambda <1$ imposed as an additional constraint in \eqref{LPf}.}
\end{thm}
\noindent \textit{Proof:}
Using \eqref{clset}, Lemma \ref{trans2} and  the  Minkowski sum of two constrained zonotopes \cite{setcont2}, one has 
\begin{align} \label{next2}
 &   x(t+1) \in \Big<\Big[ -{G}^w \circ (V^K x(t)) \quad \mathbf{0}_{n \times s_\theta} \quad G^h\Big], \nonumber \\ & \quad \quad \quad  \quad \quad (X^1-C^w) V^K x(t)+c^h,{\tilde{A}}^c,{\tilde{b}}^c\Big>,
\end{align}
where 
$
   {\tilde{A}}^c=\begin{bmatrix}
       \text{Vec}(A^C) & \mathbf{0} \\
       \mathbf{0} & \mathbf{0}
   \end{bmatrix}, \,\,
   {\tilde{b}}^c= \begin{bmatrix}
       \text{Vec}(B^C) \\ \mathbf{0}
   \end{bmatrix}.
$

By definition, $\lambda$-contractivity is satisfied if $\gamma_j \le \lambda \, {h}_j$, where $\gamma_j=\max\limits_{x(t)} \, {H}_{j,:} x(t+1)$, $j=1,...,q$, whenever $H x(t) \le h$. Using \eqref{next2}, define
\begin{align*} 
 & \bar \gamma_j  = \max\limits_{\beta} \max\limits_{x(t)}  \Big( {H}_{j,:} \, (X^1-C^w) V^K \, x(t)+{H}_{j,:} \, c^h+ \nonumber \\ &   \Big|\sum\limits_{i=1}^{Ts_w}   \beta_i {H}_{j,:} \, G^{w}_{:,i} V^K x(t) \Big|   +  \sum\limits_{i=1}^{s_w}  \big|{H}_{j,:} \,  G^h_{:,i} \big| \Big) \nonumber \\  &
 {\rm{s}}{\rm{.t}}{\rm{.}} \,\, \sum\limits_{i=1}^{s_c} {A}^{C}_{:,i} \, \beta_i={B}^{C},  \quad \big\Vert \beta  \big\Vert \leq 1, \nonumber \\  & \quad \quad H x(t) \le h,
\end{align*}
where the last two terms of the cost function are obtained using the fact that for $\mu=[\beta^{\top} \,\,\, \eta^{\top}]^\top \in \mathbb{R}^{s_c+s_w}$, one has
\begin{align} \label{ineql}
   &  {H}_{j,:} \, [-G^w \circ (V^K x(t)) \quad \mathbf{0}_{n \times s_\theta}  \quad G^h] \mu  \le  \nonumber \\ & 
     \Big|\sum\limits_{i=1}^{Ts_w}   \beta_i{H}_{j,:} \, G^{w}_{:,i} \,  V^K x(t)\Big| + \sum\limits_{i=1}^{s_{w}}   \Big|{H}_{j,:} \,  G^h_{:,i} \Big|,
\end{align}
where the second term results from $| \sum_{i=1}^{s_{w}} {H}_{j,:}  \, G^h_{:,i} \, \eta_i| \le \sum_{i=1}^{s_{w}} |{H}_{j,:}  \, G^h_{:,i}|$, using $\Vert \eta\Vert  \le 1$. Based on \eqref{ineql}, $\gamma_j \le \bar \gamma_j$, and thus, Problem 1 is solved if $\bar \gamma_j \le \lambda h_{j}$ for $j=1,...,q$. To find a bound for the term depending on $\beta_i$ in the cost function, since ${H} x(t) \leq {h}$, a multi-dimensional interval enclosure of this polytope can be obtained for $x(t)$ as $x(t) \in [\underline{x} \quad \overline{x}]$, where $\underline{x}, \overline{x} \in \mathbb{R}^n,$ and $\underline{x} \le \overline{x} $ \cite{Althoff}. From this, a bound on $\| x(t)\| $ is derived depending on $h$ and $H$ (i.e., $\|x\| \le M_x$ for some $M_x$ depending on $h$ and $H$). 
Therefore, $| \sum_{i=1}^{Ts_w} \beta_i {H}_{j,:} \, G^{w}_{:,i} \, V^K x(t)\Big| \le \| \sum_{i=1}^{Ts_w} \beta_i {H}_{j,:}  \,  {G^w_{:,i}}\Vert \Vert  {V^K} \Vert M_x$. Letting $\|V^K\| \le \rho$ and optimizing over $\rho$, one has $\bar \gamma_j \le \hat \gamma_j$, where 
\begin{align*} 
 & \hat \gamma_j  =  \max\limits_{x(t)}  \Big( {H}_{j,:} \, (X^1-C^w)  V^K \, x(t)+{H}_{j,:} \, c^h+  \nonumber \\ &  \quad \quad \quad\quad \quad\big\Vert {V^K}  \big\Vert l_j+y_j \Big) \quad
 {\rm{s}}{\rm{.t}}{\rm{.}}   \quad H x(t) \le h,
\end{align*}
with $y_j$ and $l_j$ defined in \eqref{yj} and \eqref{lj}. 
Using duality, $\lambda$-contractivity is satisfied if $\tilde \gamma_j \le \lambda {h}_j$ where 
\begin{align*} 
&\tilde \gamma_j  = \min\limits_{\alpha_j} \alpha_j^{\top} \, h+  {H}_{j,:} \, c^h+ l_j \big\Vert {V^K} \big\Vert+y_j, \label{a1} \nonumber \\&
{\alpha_j}^{\top} H =  {H}_{j,:} \, (X^1-C^w)  V^K,  \nonumber \\&
{\alpha_j}^{\top} \ge \mathbf{0}_{1 \times q},  
\end{align*}
where $\alpha_i \in \mathbb{R}^q$. Define $
{P}=[\alpha_1,....,\alpha_q]^{\top} \in \mathbb{R}^{q \times q}$. $P$ is non–negative since $\alpha_i$ is non-negative for all $i=1,...,q$. Therefore, using the dual optimization,  $\lambda$-contractivity is satisfied if \eqref{LPf} is satisfied. \hfill   $\blacksquare$ 

\textcolor{blue}{We now present Algorithm 1 to adaptively update the controller as new data become available. 
 The offline phase uses the offline data from \eqref{data-u}–\eqref{data-z} satisfying Assumption 5, while the online phase updates the controller every $T^s$ steps without any assumption on data.
In \eqref{LPf}, when $r \!=\! \lambda$, Algorithm 1 optimizes over $\lambda$ to adaptively improve performance. In contrast, $r \!=\! \rho$ uses a fixed $\lambda$ to highlight disturbance tolerance.}

\begin{algorithm}
\textcolor{blue}{\caption{\textbf{Adaptive Unified Learning for Safe Control}} \label{Alg:explorexploitAlg}
\begin{algorithmic}[1]
    \STATE \textbf{Inputs:} $\cal{M}_{prior}$, $\cal{Z}_w$, $\cal{S}_s$, $X^1$, $X^0$, $U^0$.
    \STATE \textbf{Offline Learning:}
    \STATE Use Theorem 1 to characterize the CMZ set of closed-loop dynamics. Design a safe controller via Theorem 2.
    \STATE \textbf{Online Learning and Adaptation:}
        \FOR{$t = 1,...,T^s$}
                \STATE Apply the learned controller to the system. 
         \ENDFOR
                \STATE Check the informativeness of the new $T^s$ data samples using Lemma 3.
                \STATE If informative, use Lemma 4 and 5 to replace $\cal{M}_{prior}$ with the updated FOLMS.
                \STATE Compute the new controller using the updated $\cal{M}_{prior}$ in Theorem 2.
       \STATE Move to Step 5.    
\end{algorithmic}}
\end{algorithm}

\begin{figure*}[!t]
    \centering
    \begin{minipage}{0.33\textwidth}
        \centering
        \includegraphics[width=\textwidth]{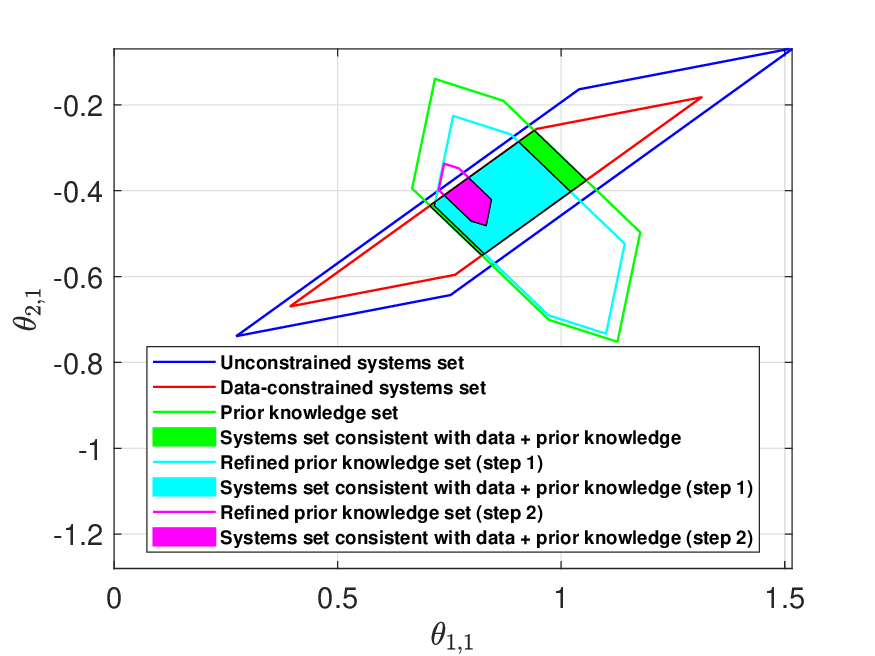}
        \subcaption{} 
        \label{l1a}
    \end{minipage}%
    \begin{minipage}{0.33\textwidth}
        \centering
        \includegraphics[width=\textwidth]{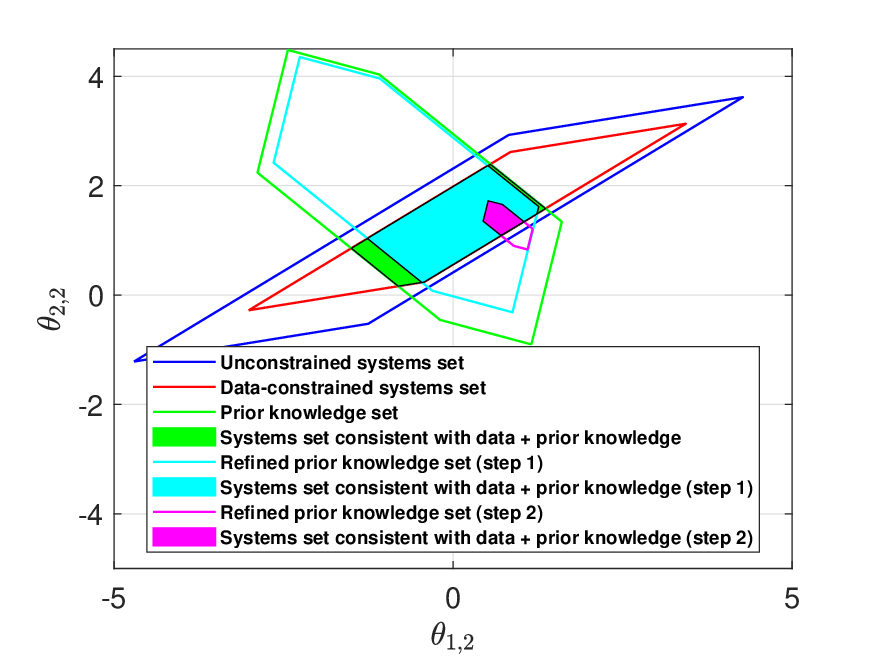}
        \subcaption{} 
        \label{l1b}
    \end{minipage}
    \begin{minipage}{0.33\textwidth}
        \centering
        \includegraphics[width=\textwidth]{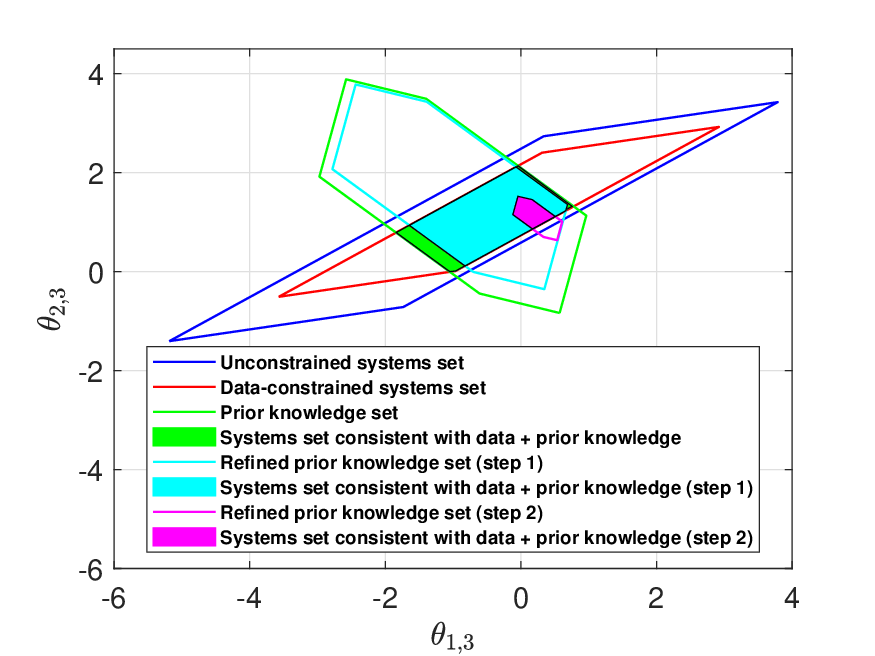}
        \subcaption{} 
        \label{l1c}
    \end{minipage}%
    \caption{\small{\textcolor{blue}{The dark blue region shows the unconstrained set derived from data. The red region incorporates only $\bar{A}^w$ and $\bar{B}^w$ in $A^C$ and $B^C$. The green region further applies prior knowledge constraints, which may not align with the data due to many reasons (e.g., loose empirical priors, unmodeled disturbances, and regime shifts). Across all plots, the initial filled green region represents the first uncertainty set, reducing the unconstrained set by approximately $73-77\%$. Prior refinement yields a smaller set (light blue), with a $76-81\%$ reduction, and further updates lead to the final refined set (purple), achieving a total reduction of $97-98\%$ relative to the original unconstrained system set.}}}
    \label{l1}
\end{figure*}


\textcolor{blue}{Given Step 9 in Algorithm~1, the size of the set of closed-loop systems is monotonically decreasing, ensuring that control adaptation progressively improves performance without compromising feasibility. The convergence analysis of our online set-membership identification approach can be framed using the results in~\cite{Converg}. Similar to \cite{Converg}, which considers disturbances bounded by convex sets, one can show that Algorithm 1 achieves a convergence rate of $\mathcal{O}(\frac{1}{\mathcal NT^s})$, where $\mathcal N$ is the number of times Algorithm 1 adapts the controller based on $T^s$ online samples. 
However, as demonstrated both in our simulations and those of~\cite{Converg}, a discrepancy exists between theoretical predictions and empirical behavior—primarily due to the assumptions on the disturbance distribution and the challenge of collecting rich data. The former challenge stems from the fundamental difficulty of satisfying pointwise boundary-visiting disturbances. The latter arises from the need to sufficiently excite the system (i.e., explore) to collect rich, independent data—without compromising safety during the process. We leave this challenge for future work. In practice, the set of closed-loop systems shrinks during the online learning process and converges to a limit determined by the actual size of the disturbance set.}
\textcolor{blue}{\begin{rem}
Traditional robust control assumes a controller exists that stabilizes all models within the prior knowledge set. In contrast, Theorem~\ref{th3} relies on closed-loop models informed by both data and prior knowledge. Thus, a solution requires a control gain that stabilizes all dynamics in this smaller, refined set. If the prior and disturbance sets are large and the data is insufficient, a solution may not exist—no controller can be expected to succeed under such uncertainty. Another factor influencing the existence of a solution is the center of the disturbance. If \( c^h \neq 0 \) and a solution exists, safety can still be certified; however, the system trajectories asymptotically approach a bounded neighborhood of a nonzero equilibrium point. To eliminate the offset, the control structure can be revised (e.g., by adding an integrator) to reject \( c^h \).
\end{rem}}\vspace{-10pt}
\textcolor{blue}{\begin{rem}
The linear program \eqref{LPf} is solved efficiently, as the size of the decision variable $V^K$ depends solely on the number of offline samples and remains independent of the number of online samples.
With abundant offline data, the closed-loop representation in~\cite{NewG} reduces the decision variables to depend only on system dimensions, a setting to which our results extend. Besides, when the safe set is non-convex, convex lifting decomposes it into a finite set of polyhedral \cite{convexlift}. 
\end{rem}}

\section{Simulation Results}

Consider the system in~\eqref{system}, where the true but unknown dynamics $A^* \in \mathbb{R}^{2 \times 2}$ and $B^* \in \mathbb{R}^2$, are given as in~\cite{example}.
Let the additive disturbance $w(t)$ lies within a zonotope $\cal{Z}_w\!=\!\big<\!G^h,c^h\!\big>$, with 
\(
G^h\!=\!\alpha\begin{bmatrix}
    0.05 & 0.08 \\
    0.01 & 0.06
\end{bmatrix}\), and \(c^h \!=\! \begin{bmatrix}
    0 \\
    0
\end{bmatrix}, 
\)
where $\alpha$ is the noise scaling factor.
The safe set in \cite{example} is considered as a polytope $\cal{S}_s=\{x: Hx \le h \}$, where $h=[1,1,1,1]^{\top}$ and $H$ is defined in \cite{example}.
The prior knowledge is modeled as an MZ, where its center $C^\theta$ and generator $G^\theta$ are formed using 5 data obtained by applying stabilizing input to the system in \eqref{system}, but under a different noise zonotope set $\big <\!\begin{bmatrix}
    0.06 & -0.12 & 0.02\\
    -0.02 & 0.12 & 0.1
\end{bmatrix}, \begin{bmatrix}
    1 \\
    -1
\end{bmatrix}\!\big>$. Unless otherwise noted,  simulations use $T \!=\! 5$, $T^s\!=\! 1$, $\alpha \!=\! 1$, and $r \!=\! \rho$ with $\lambda \!=\! 0.99$.
\begin{figure}[t]
    \centering
    \includegraphics[width=0.35\textwidth, keepaspectratio]{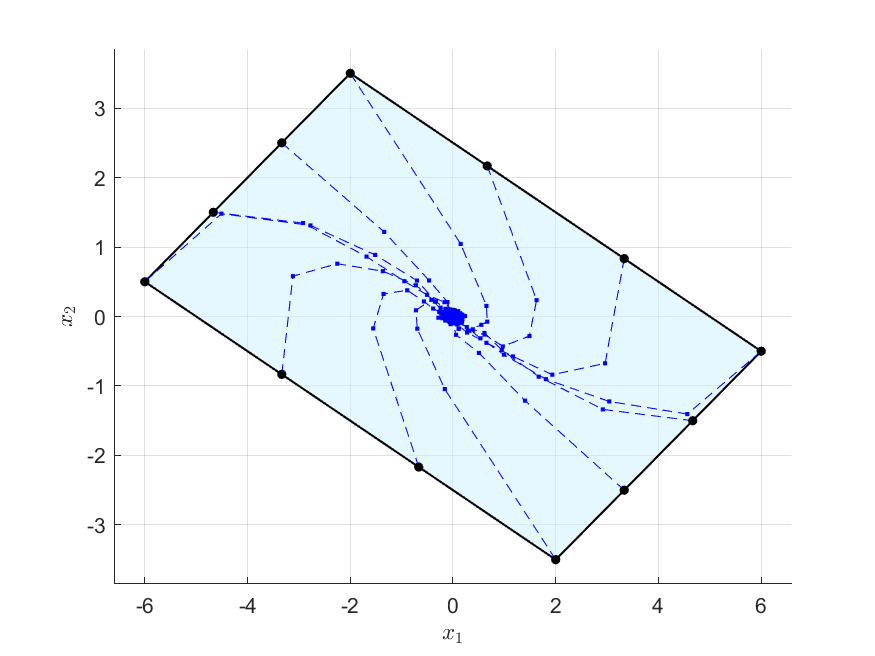} 
    \caption{\small{Phase portrait of the trajectories \textcolor{blue}{from a single run}, originating from different initial points on the boundary of the safe set.}}
    \label{phase}
\end{figure}

A stabilizing input is first applied to the system to collect data. Fig.~\ref{l1} illustrates the system set \( \theta = [A \;\; B] \in \mathbb{R}^{2 \times 3} \) in all dimensions, with $r=\rho$ in \eqref{LPf} in Algorithm 1. As shown in Fig. 1, incorporating only \( \bar{A}^w \) and \( \bar{B}^w \) into \( A^C \) and \( B^C \) (as in Theorem~\ref{clrep}) yields the red region, reducing the initial uncertainty shown by dark blue. Adding prior knowledge constraints further shrinks the set, forming the green region. The closed-loop matrix zonotope \( \cal{M}_{cl} \) is then constructed via~\eqref{clset}, as detailed in Theorem~\ref{clrep}.
After collecting a few online data samples, the uncertainty set tightens (blue region), and is further refined through successive refinement steps (e.g., the purple region). 

To show the performance improvement, we set $r=\lambda$ in \eqref{LPf} when using Algorithm 1. \textcolor{blue}{For $T = 15$, Monte Carlo simulations over 100 trials show that Algorithm 1 decreased $\lambda$ on average by 0.16. Among the trials, the largest observed decrease was 0.31 (from 0.83 to 0.52), clearly indicating improved performance as more data becomes available.}
As shown in Fig.~\ref{phase}, the phase portrait shows that all trajectories remain within the safe region and \textcolor{blue}{converge to a neighborhood of the origin}, thereby guaranteeing both safety and stability.  

\begin{figure}[ht]
  \centering
  \begin{subfigure}{0.49\linewidth}
    \includegraphics[width=\linewidth, height=1.6in]{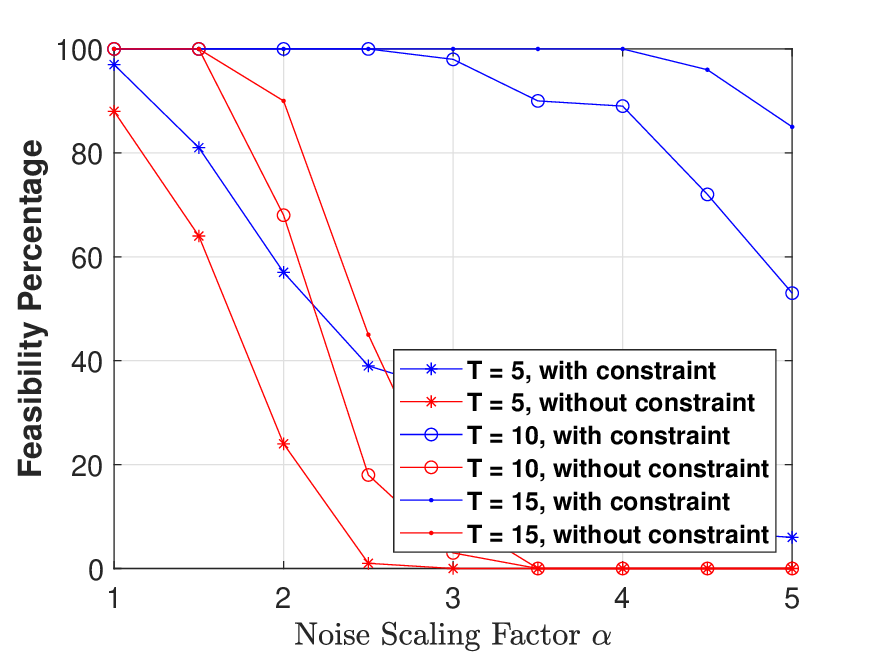}
    \caption{}
    \label{alpha}
  \end{subfigure}%
  \hspace{0.01\linewidth}
  \begin{subfigure}{0.49\linewidth}
    \includegraphics[width=\linewidth, height=1.6in]{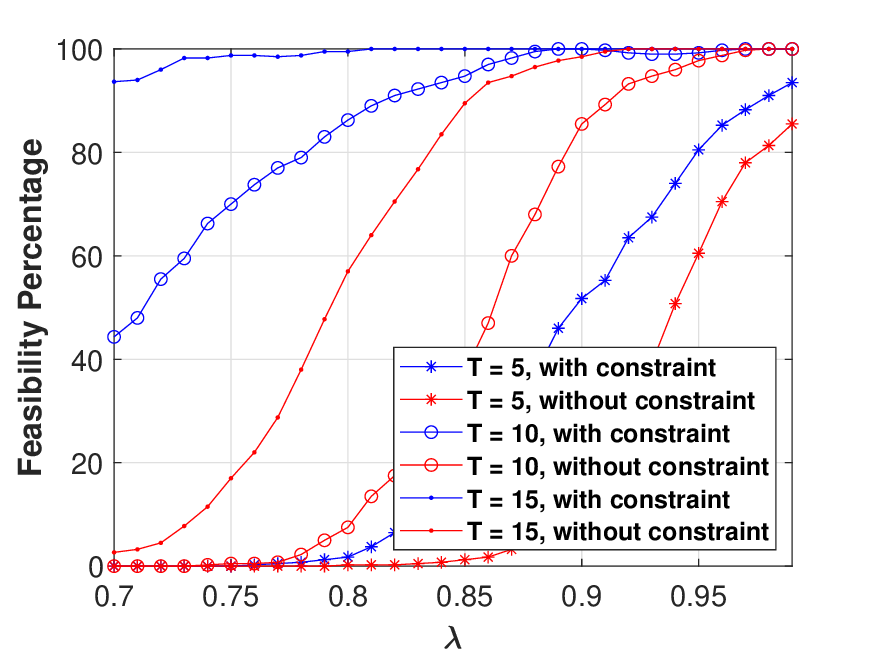}
    \caption{}
    \label{lambda}
  \end{subfigure}
  \caption{\small{Feasibility rate of the optimization problem \eqref{LPf} as a function of (a) the scaling factor $\alpha$ and (b) the contraction factor \( \lambda \), averaged over 100 Monte Carlo simulations.}}
  \label{Alpha-Lambda}
\end{figure}

To assess the impact of uncertainty modeling on the feasibility and conservativeness of the controller synthesis, we compare two cases in the computation of \( l_j \) in \eqref{lj}: (i) with the constraint \eqref{Eq1} is enforced (i.e., when data and prior knowledge are taken into account), and (ii) without enforcing it. 
  Fig.~\ref{alpha} shows the feasibility percentage of the optimization problem \eqref{LPf} in terms of the noise scaling factor with $\lambda = 0.99$. As depicted, our conformal constrained approach exhibits a higher feasibility rate for all values of $T$. As the data length increases, the uncertainty set becomes smaller, resulting in improved feasibility. 
  
  Moreover, as \( \lambda \) decreases, \( \lambda \cal{P} \) tightens, requiring the controller to better counter disturbances. This makes satisfying contractivity harder, especially with conservative disturbance estimates (via \( l_j \)) is overly conservative. Fig.~\ref{lambda} illustrates the feasibility percentage of the optimization problem \eqref{LPf} as a function of the contraction factor \(\lambda\), for \(\alpha = 1\). The results show that although decreasing \(\lambda\) reduces the feasibility rate, our conformal constrained approach consistently achieves a higher feasibility rate for each value of \(T\). Moreover, increasing \(T\) improves feasibility; notably, for \(T = 15\),  the feasibility rate remains above 90\% even for \(\lambda = 0.7\). This highlights that incorporating the data and prior knowledge equality constraint allows the optimization problem to remain feasible under tighter contractivity requirements, thereby enabling the controller to maintain robustness against higher-level disturbances.

\textcolor{blue}{Simulations are performed in \textsc{MATLAB} on an Intel Core i7-10700 CPU @ 2.90\,GHz and 32\,GB RAM. For $T = 15$, the average computation time over 100 runs is 0.15 seconds.}

\section{conclusion} A novel approach integrates prior knowledge and open-loop learning with closed-loop learning for safe control design of linear systems. A data-based closed-loop characterization is presented, and conformal equality constraints are imposed to limit the set of closed-loop systems to those that can be explained by data and prior knowledge. The prior knowledge itself is refined by learning a zonotope-based set membership identification. We then leveraged the $\lambda$-contractivity to impose constrained zonotope safety. Future work will extend these results to nonlinear systems \textcolor{blue}{using polynomial zonotopes.}

\bibliographystyle{IEEEtran}

\end{document}